\documentclass[a4paper,11pt]{article}

\usepackage[margin=2.35cm]{geometry}
\usepackage[T1]{fontenc}
\usepackage[utf8]{inputenc}
\usepackage{lmodern}
\usepackage{amsmath,amssymb,mathtools,bm}
\usepackage{graphicx}
\usepackage{subcaption}
\usepackage{booktabs}
\usepackage{array}
\usepackage{multirow}
\usepackage{enumitem}
\usepackage{xcolor}
\usepackage[numbers,sort&compress]{natbib}
\usepackage{hyperref}
\hypersetup{colorlinks=true,citecolor=blue,linkcolor=blue,urlcolor=blue}
\graphicspath{{figures/}}
\allowdisplaybreaks

\newcommand{\dd}{\mathrm{d}}

\newcommand{\GeV}{\mathrm{GeV}}

\newcommand{\MM}{\mathrm{MM}}

\title{Neural solutions of coupled ghost and gluon Dyson--Schwinger equations in Landau gauge}
\author{Rodrigo Carmo Terin\\[2mm]
\small King Juan Carlos University, Faculty of Experimental Sciences and Technology,\\
\small Department of Applied Physics, Av. del Alcalde de M\'ostoles, 28933 Madrid, Spain\\
\small \texttt{rodrigo.carmo@urjc.es}}
\date{}

\begin{document}
\maketitle

\begin{abstract} 
The coupled ghost and gluon Dyson--Schwinger equations (DSEs) of four-dimensional Landau-gauge Yang--Mills (YM) theory are solved with a neural representation trained only from renormalized equation residuals. The neural and fixed-point solutions agree at the percent level and remain stable under changes of initialization, network size, integration grid, and infrared boundary condition. Variations of the three-gluon vertex model produce substantially larger effects than the neural error. The MiniMOM ultraviolet running and the sign change of the gluon Schwinger function are also reproduced within the limitations of the truncation. 
\end{abstract} 


\section{Introduction}
\label{sec:introduction}

Quantum chromodynamics (QCD) describes the strong interaction between quarks and
gluons~\cite{MarcianoPagels1978,Brambilla2014}.  At large momentum scales its
coupling becomes weak and perturbation theory can be used
\cite{GrossWilczek1973,Politzer1973}.  At hadronic scales, however, the theory
has to be treated nonperturbatively.  Confinement and dynamical chiral symmetry
breaking are two fundamental properties in this regime.  Their description is
closely related to the correlation functions of the theory
\cite{RobertsWilliams1994,AlkoferSmekal2001,RobertsSchmidt2000,Fischer2019}.
These correlation functions also enter continuum calculations of hadrons and
their properties~\cite{Bashir2012,Eichmann2016,DingRobertsSchmidt2023}.
Different methods are available for the nonperturbative regime, e.g., Lattice
calculations furnish a first-principles formulation on discretized space-time
and have been used successfully for many quantities~\cite{Wilson1974,Aoki2017}.
Functional equations constitute a continuum alternative. DSEs, functional renormalization-group equations, and $n$PI equations can
be derived from the gauge-fixed action however in practical computations, the
infinite systems have to be truncated
\cite{Dyson1949,Schwinger1951a,Schwinger1951b,RobertsWilliams1994,Huber2020Review}.
Consequently, two different problems have to be considered:  the retained
equations must be solved with sufficient numerical accuracy, and the effect of
the neglected or modeled correlation functions must be understood.

During the last years, larger systems of functional equations have become
accessible.  Model input was replaced by dynamically calculated vertices and
the effect of additional diagrams was investigated
\cite{Huber2020Review,Huber2020Corr}.  This progress led, among other results,
to a parameter-free calculation of Landau-gauge correlation functions and to
a glueball spectrum in quantitative agreement with lattice simulations 
\cite{Huber2020Corr,HuberFischerSanchis2020}.  It should be noted that some
aspects which initially appear to be only technical become important at this
level.  Examples are spurious divergences in the gluon equation and the
realization of the perturbative resummation in a truncated system
\cite{HuberSmekal2014,Huber2020Review}.
The Landau-gauge YM propagators provide a simple system in which these
issues can be studied.  The ghost and gluon equations are coupled nonlinear
integral equations.  In four dimensions they have to describe the infrared
solution and the logarithmic ultraviolet running at the same time.  The
infrared solutions form a family of decoupling solutions with an endpoint
scaling solution, whereas four-dimensional lattice calculations yield the
decoupling type~\cite{FischerMaasPawlowski2009,Huber2020Review}.  Quantitative
results also depend on the three-gluon vertex.  In reduced propagator
truncations this vertex is normally modeled, and different choices can change
the gluon propagator considerably
\cite{HuberSmekal2013,Blum2014,Eichmann2014}.

Neural representations offer another way to solve differential and integral
equations.  Early approaches used trial functions in which boundary
conditions were incorporated explicitly~\cite{Lagaris1998}.  In
physics-informed neural networks (PINNs) the equation residual is part of the loss
function~\cite{Raissi2019,Karniadakis2021}.  Related techniques were applied to
many-body wave functions, quantum-mechanical systems, and operator learning
\cite{Pfau2020,Brevi2024,Lu2021}.  For functional equations, a neural
representation is attractive because it is continuous and differentiable and
can be inserted directly into nonlocal integrals.  On the other hand, a small
loss on the training points is not sufficient.  The residual must also be
checked with an independent quadrature, and observables obtained from the
solution may require a distinct convergence analysis.
Physics-informed neural solutions of Euclidean QED Dyson--Schwinger equations
were studied in our one of previous works.~\cite{Terin2025SciPost}.  Additionally, our recent Minkowski-space analysis
showed that constraints which are useful in one regime can become inconsistent
when the analytic structure changes~\cite{Terin2026JHEP}.  Here the method is
applied to the YM ghost and gluon propagators.  A reduced one-loop
truncation is used deliberately.  The same equations are solved with a direct
fixed-point iteration and with a neural fixed-point procedure.  Our purpose is
to determine whether the neural representation reaches the same solution
without propagator data and to compare its numerical error with discretization
and vertex-model effects.  The ultraviolet behavior and the Schwinger
function are considered as further tests.

The coupled equations and their renormalization are introduced in Sec.~\ref{sec:equations}. The direct and neural solution methods are described in Sec.~\ref{sec:methods}. Numerical results, including the convergence tests and the dependence on the three-gluon vertex, are presented in Sec.~\ref{sec:results}. The results are discussed in Sec.~\ref{sec:discussion}, and Sec.~\ref{sec:conclusions} contains the conclusions. Additional neural convergence tests are collected in Appendix~\ref{app:neural-tests}, while the physical and numerical parameters used in the calculations are summarized in Appendix~\ref{app:parameters}.

\section{Coupled ghost and gluon equations}
\label{sec:equations}

We consider pure $SU(N_c)$ YM theory in four-dimensional Euclidean
space.  The Landau gauge is fixed by the condition $\partial_\mu A_\mu^a=0$.
The gluon propagator is then transverse~\cite{FaddeevPopov1967}.  The ghost and
gluon propagators are written as
\begin{align}
 D_G^{ab}(p)&=-\delta^{ab}\frac{G(p^2)}{p^2},
 \label{eq:ghost-propagator}\\
 D_{\mu\nu}^{ab}(p)&=\delta^{ab}P_{\mu\nu}(p)\frac{Z(p^2)}{p^2},
 \qquad
 P_{\mu\nu}(p)=\delta_{\mu\nu}-\frac{p_\mu p_\nu}{p^2}.
 \label{eq:gluon-propagator}
\end{align}
The scalar part of the gluon propagator is denoted by
\begin{equation}
 D(p^2)=\frac{Z(p^2)}{p^2}.
 \label{eq:scalar-gluon}
\end{equation}
The solutions considered below are of the decoupling type.  Thus, the ghost
dressing is finite in the infrared and the gluon propagator approaches a
nonzero value.  It should be noted that positive dressing functions on the
Euclidean momentum axis do not imply a positive spectral representation.
The propagator equations are truncated at one loop and the ghost equation
contains the ghost--gluon loop.  In the gluon equation, the ghost and gluon
loops are retained.  The ghost--gluon vertex is kept at tree level and for the
three-gluon vertex, the tree-level tensor is multiplied by a scalar dressing.
The tadpole is absorbed into the infrared condition and the sunset and squint
diagrams are not included.  This system is a reduced version of the
truncations discussed in Refs.~\cite{vonSmekal1998,HuberSmekal2013,Huber2020Review}.
Its size is still sufficient to test a coupled neural solution, while the same
integral equations can be solved independently by a conventional iteration.
With $x=p^2$, $y=q^2$, and $z=(p+q)^2$, the scalar equations read
\begin{align}
 \frac{1}{G(x)}
 &=\widetilde Z_3+N_c g^2\int_q
 Z(y)G(z)K_G(x,y,z),
 \label{eq:ghost-dse}\\
 \frac{1}{Z(x)}
 &=Z_3+N_c g^2\int_q
 \left[
 G(y)G(z)K_Z^{\rm gh}(x,y,z)
 +Z(y)Z(z)K_Z^{\rm gl}(x,y,z)C^{AAA}(x,y,z)
 \right]-\frac{C_{\rm sub}}{x},
 \label{eq:gluon-dse}
\end{align}
where $\int_q=\int\dd^4q/(2\pi)^4$.  The angle between the external and loop
momenta is denoted by $u=\cos\theta$.  With the transverse projector, the
kernels are
\begin{align}
 K_G(x,y,z)&=-\frac{1-u^2}{yz},
 \label{eq:kernel-ghost}\\
 K_Z^{\rm gh}(x,y,z)&=\frac{1-u^2}{3xz},
 \label{eq:kernel-gluon-ghost}\\
 K_Z^{\rm gl}(x,y,z)
 &=-\frac{2(1-u^2)}{3xyz^2}
 \Big[u^2xy+6u\sqrt{xy}(x+y)+3x^2+8xy+3y^2\Big].
 \label{eq:kernel-gluon-gluon}
\end{align}
These expressions are equivalent to the one-loop kernels collected in
Ref.~\cite{Huber2020Review}.
For the baseline calculation, the three-gluon-vertex dressing is chosen as
\begin{equation}
 C^{AAA}_{\rm base}(x,y,z)
 =\frac{G(\bar p^2)}{Z(\bar p^2)}
 \frac{\bar p^2}{\bar p^2+\Lambda_s^2},
 \qquad
 \bar p^2=\frac{x+y+z}{2},
 \label{eq:three-gluon-model}
\end{equation}
with $\Lambda_s^2=1.54\,\GeV^2$.  The factor $G/Z$ implements a simple
renormalization-group improvement.  Without an additional infrared damping,
this form leads to a strong enhancement in the gluon loop.  The rational
factor is therefore included.  Similar effective constructions were used in
propagator calculations~\cite{HuberSmekal2013,Huber2020Review}.  They should
not be confused with a dynamically calculated three-gluon vertex.
In order to estimate the effect of this input, three further choices are used:
\begin{align}
 C^{AAA}_{\rm bare}&=1,
 \label{eq:vertex-bare}\\
 C^{AAA}_{\rm supp}&=\frac{\bar p^2}{\bar p^2+\Lambda_s^2},
 \label{eq:vertex-suppression}\\
 C^{AAA}_{\rm no\,IR}&=\frac{G(\bar p^2)}{Z(\bar p^2)}.
 \label{eq:vertex-noir}
\end{align}
The bare and suppression-only models remove the $G/Z$ factor in different
ways.  The last choice keeps this factor without infrared suppression and is
used to test the stability of the truncated equations.

The ghost equation is renormalized by a momentum subtraction.  When a hard
cutoff is used, the gluon equation also contains a spurious quadratic
divergence.  Different prescriptions for its subtraction are known
\cite{HuberSmekal2014,Huber2020Review}.  Here a mass counterterm is fixed by an
infrared propagator condition.  The equations are specified by
\begin{equation}
 G(x_m)=G_m,
 \qquad
 Z(x_s)=1,
 \qquad
 D(x_m)=D_0,
 \label{eq:renormalization-conditions}
\end{equation}
where $x_m$ is an infrared point and $x_s$ is the subtraction point.
Let $\Sigma_G$ and $\Sigma_Z$ denote the loop contributions in the ghost and
gluon equations.  The subtracted ghost equation is
\begin{equation}
 \frac{1}{G(x)}
 =\frac{1}{G_m}+\Sigma_G(x)-\Sigma_G(x_m).
 \label{eq:ghost-subtracted}
\end{equation}
The two gluon conditions give
\begin{equation}
 C_{\rm sub}
 =\frac{x_mx_s}{x_s-x_m}
 \left[\Sigma_Z(x_m)-\Sigma_Z(x_s)\right]
 +\frac{x_mx_s}{x_s-x_m}
 -\frac{x_s}{x_s-x_m}D_0^{-1}.
 \label{eq:csub}
\end{equation}
For the numerical solution, it is useful to write the gluon equation for the
inverse scalar propagator $\Gamma_D(x)=x/Z(x)$,
\begin{equation}
 \Gamma_D(x)
 =x+x\left[\Sigma_Z(x)-\Sigma_Z(x_s)\right]
 +C_{\rm sub}\left(\frac{x}{x_s}-1\right).
 \label{eq:inverse-propagator-subtracted}
\end{equation}
This form avoids a cancellation of terms proportional to $1/x$ in the
infrared.  It should be noted that the value of $C_{\rm sub}$ depends on the
regulator and on the quadrature.  Only the renormalized propagator is compared
below.
The MiniMOM coupling is defined by
\begin{equation}
 \alpha_{\MM}(p^2)
 =\alpha(x_s)G^2(p^2)Z(p^2),
 \label{eq:minimom-coupling}
\end{equation}
where the finiteness of the Landau-gauge ghost--gluon vertex in Taylor
kinematics is used~\cite{vonSmekal1997,vonSmekal2009,Boucaud2009}.

\section{Numerical methods}
\label{sec:methods}

The standard parameters are
\begin{equation}
 \alpha(x_s)=0.05,
 \quad
 G_m=10,
 \quad
 D_0=15.54\,\GeV^{-2},
 \quad
 x_m=10^{-5}\,\GeV^2,
 \quad
 x_s=7720\,\GeV^2,
 \label{eq:baseline-parameters}
\end{equation}
with $N_c=3$.  External momenta are taken from
$10^{-5}$ to $10^6\,\GeV^2$.  The radial integration extends from
$10^{-8}$ to $10^8\,\GeV^2$.  After the transformation $v=\ln y$, the radial
integral is evaluated with Gauss--Legendre quadrature.  For the angular
integral,
\begin{equation}
 \int_{-1}^{1}\dd u\,\sqrt{1-u^2}\,f(u),
\end{equation}
Gauss--Chebyshev quadrature of the second kind is used.  The initial grid
contains $140$ external points, $140$ radial nodes, and $36$ angular nodes.
The equations are solved by an under-relaxed fixed-point iteration and logarithmic
variables are used to keep the dressing functions positive.  If the
right-hand sides at iteration $n$ give $G_{\rm DSE}^{(n)}$ and
$Z_{\rm DSE}^{(n)}$, the update is
\begin{align}
 \ln G^{(n+1)}&=(1-\omega)\ln G^{(n)}+\omega\ln G_{\rm DSE}^{(n)},\\
 \ln Z^{(n+1)}&=(1-\omega)\ln Z^{(n)}+\omega\ln Z_{\rm DSE}^{(n)},
\end{align}
with $\omega=0.08$.  The iteration is stopped when the largest logarithmic
change is smaller than $2\times10^{-6}$.
The input of the network is $t=\ln x$.  It has two outputs, $N_G(t)$ and
$N_Z(t)$.  These functions multiply analytic baseline functions which satisfy
the renormalization conditions.  The baseline network has two hidden layers
of width $40$ and hyperbolic-tangent activation functions.  Double precision is
used throughout.  The implementation is based on PyTorch~\cite{Paszke2019}.
For the ghost dressing, the standard function is
\begin{equation}
 G_{\rm base}(x)
 =1+(G_m-1)
 \frac{1+(x_m/s_G)^{\nu_G}}{1+(x/s_G)^{\nu_G}},
 \qquad
 s_G=0.1\,\GeV^2,
 \qquad
 \nu_G=\frac12,
 \label{eq:ghost-baseline}
\end{equation}
and $G_{\rm base}(x_m)=G_m$.  For the gluon dressing, we use
\begin{equation}
 Z_{\rm base}(x)
 =\frac{x/(x+m_0^2)}{x_s/(x_s+m_0^2)},
 \qquad
 m_0^2=\frac{x_s(1-D_0x_m)}{D_0x_s-1},
 \label{eq:gluon-baseline}
\end{equation}
which gives $Z_{\rm base}(x_s)=1$ and
$Z_{\rm base}(x_m)/x_m=D_0$.
The neural dressing functions are defined by
\begin{align}
 G_\theta(x)
 &=G_{\rm base}(x)
 \exp\left[h_G(t)N_G(t)\right],
 \label{eq:neural-ghost}\\
 Z_\theta(x)
 &=Z_{\rm base}(x)
 \exp\left[h_Z(t)N_Z(t)\right],
 \label{eq:neural-gluon}
\end{align}
with
\begin{align}
 h_G(t)&=\tanh\frac{t-t_m}{2},\\
 h_Z(t)&=\tanh\frac{t-t_m}{2}\tanh\frac{t-t_s}{2},
\end{align}
where $t_m=\ln x_m$ and $t_s=\ln x_s$.  In this way, the three conditions in
Eq.~\eqref{eq:renormalization-conditions} are fulfilled exactly.  The
exponential form keeps the Euclidean dressing functions positive.  As noted
above, this does not impose reflection positivity or positivity of a
K\"all\'en--Lehmann density.
A direct minimization of the equation residuals from a generic initialization
turns out to be poorly conditioned.  The reason is the cancellation of large
terms in the gluon equation.  Therefore, a neural form of the fixed-point
iteration is used first.  At outer iteration $n$, the neural functions are
inserted into the right-hand sides of
Eqs.~\eqref{eq:ghost-subtracted} and
\eqref{eq:inverse-propagator-subtracted}.  This gives the functions
$G_{\rm DSE}^{(n)}$ and $Z_{\rm DSE}^{(n)}$.  Mixed targets are then defined by
\begin{align}
 \ln G_{\rm tar}^{(n)}
 &=(1-\rho)\ln G_\theta^{(n)}
 +\rho\ln G_{\rm DSE}^{(n)},\\
 \ln Z_{\rm tar}^{(n)}
 &=(1-\rho)\ln Z_\theta^{(n)}
 +\rho\ln Z_{\rm DSE}^{(n)},
 \label{eq:neural-picard-targets}
\end{align}
with $\rho=0.10$.  At every outer iteration, $50$ Adam steps are used to project
the network onto these targets.  The reference setup calculation contains $70$ outer
iterations and the targets are generated by the equations themselves.  A
converged propagator solution is not used during training.
After this preconditioning, the equation residuals are minimized directly.
Let $R_G$ and $R_D$ denote the right-hand sides of the subtracted ghost and
inverse-propagator equations.  The normalized residuals are
\begin{align}
 r_G(x)
 &=\frac{G_\theta^{-1}(x)-R_G(x)}
 {1+\tfrac12\left(|G_\theta^{-1}(x)|+|R_G(x)|\right)},
 \label{eq:normalized-ghost-residual}\\
 r_D(x)
 &=\frac{x/Z_\theta(x)-R_D(x)}
 {1+\tfrac12\left(|x/Z_\theta(x)|+|R_D(x)|\right)},
 \label{eq:normalized-gluon-residual}
\end{align}
and the loss is
\begin{equation}
 \mathcal L_{\rm DSE}
 =\left\langle r_G^2\right\rangle_x
 +\left\langle r_D^2\right\rangle_x.
 \label{eq:neural-loss}
\end{equation}
The final optimization consists of 60 Adam steps with learning rate
$3\times10^{-4}$.  The training grid has $80$ external points, 72 radial nodes,
and $18$ angular nodes.  For the validation, $120$ external points, $112$ radial
nodes, and $30$ angular nodes are used.  Thus, the residuals quoted below are
not evaluated with the training quadrature.
The direct solution is only used after the neural calculation with the purpose
to give an independent comparison.  Relative errors are defined by
\begin{equation}
 \varepsilon_2[F]
 =\frac{\|F_\theta-F_{\rm dir}\|_2}{\|F_{\rm dir}\|_2}.
 \label{eq:relative-error}
\end{equation}
For the ultraviolet analysis, the dressing functions are fitted with
\begin{equation}
 G(p^2)\propto\left[\ln\frac{p^2+\Lambda^2}{\Lambda^2}\right]^\delta,
 \qquad
 Z(p^2)\propto\left[\ln\frac{p^2+\Lambda^2}{\Lambda^2}\right]^\gamma,
 \label{eq:uv-fit}
\end{equation}
where $\Lambda^2=0.04\,\GeV^2$.  In one-loop pure YM theory,
$\delta=-9/44$, $\gamma=-13/22$, and $2\delta+\gamma=-1$ for the MiniMOM
coupling.
The zero-spatial-momentum Schwinger function is
\begin{equation}
 \Delta(t)
 =\frac{1}{\pi}\int_0^{p_{\max}}\dd p\,\cos(pt)D(p^2).
 \label{eq:schwinger-function}
\end{equation}
A negative part of this function is sufficient to establish a violation of
reflection positivity~\cite{OsterwalderSchrader1973,OsterwalderSchrader1975,Huber2020Review}.
This criterion concerns the gluon two-point function, however, it is not a complete
criterion for confinement.

\section{Numerical results}
\label{sec:results}

The direct reference setup calculation converges after $165$ iterations.  The
conditions at the infrared point are obtained as
\begin{equation}
 G(x_m)=9.9999999,
 \qquad
 D(x_m)=15.5400000\,\GeV^{-2}.
\end{equation}
The gluon dressing reaches its maximum, $Z_{\max}=2.954$, at
$p^2=0.806\,\GeV^2$.  The gluon propagator is finite in the infrared and has a
shallow maximum of $15.714\,\GeV^{-2}$ at
$p^2=8.47\times10^{-3}\,\GeV^2$.  The MiniMOM coupling goes to zero at the
infrared endpoint of this decoupling solution.  Its maximum is
$\alpha_{\MM}^{\max}=2.001$ at $p^2=0.188\,\GeV^2$.
\begin{figure}
\centering
\begin{subfigure}{0.48\textwidth}
 \includegraphics[width=\linewidth]{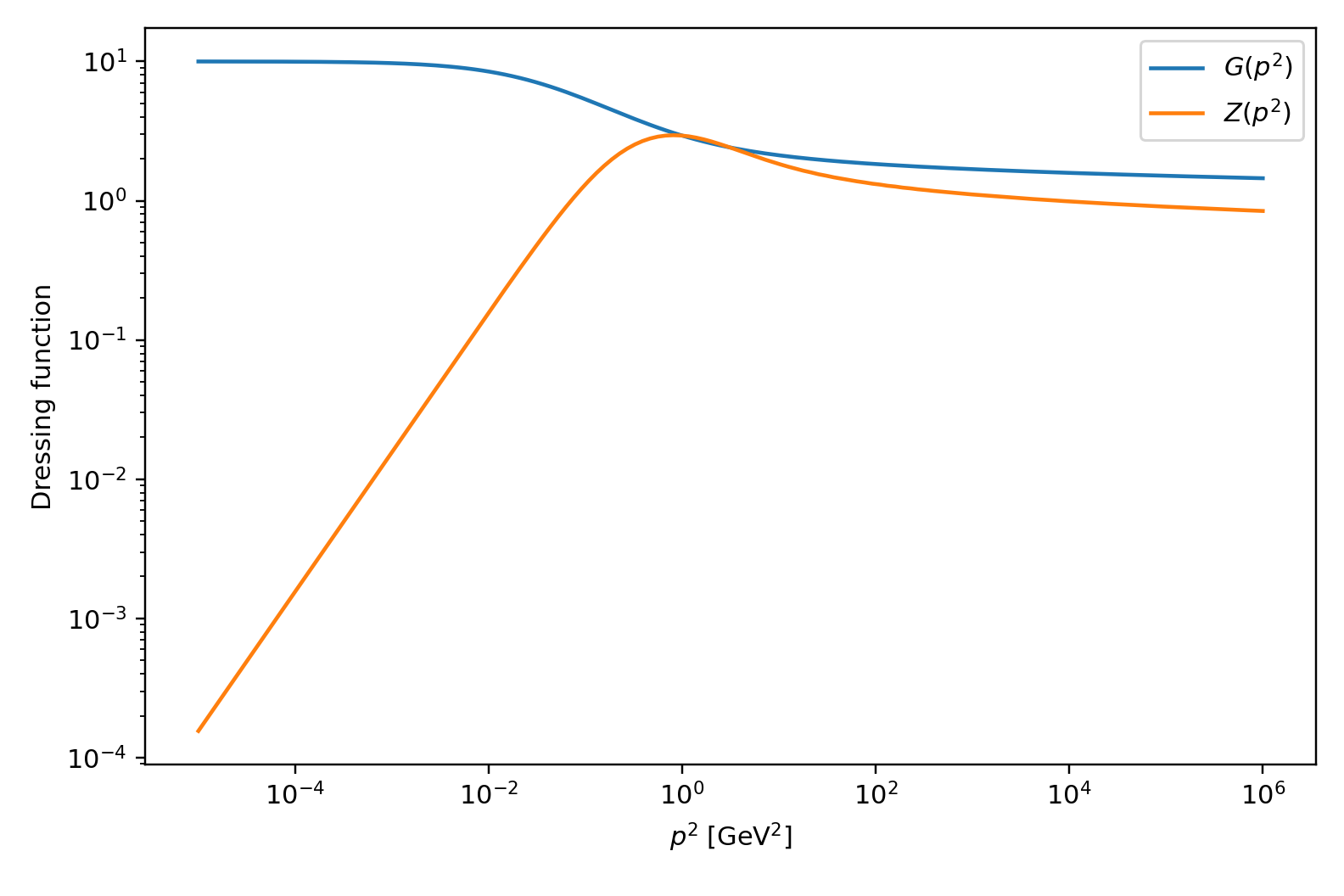}
 \caption{Ghost and gluon dressing functions.}
\end{subfigure}
\hfill
\begin{subfigure}{0.48\textwidth}
 \includegraphics[width=\linewidth]{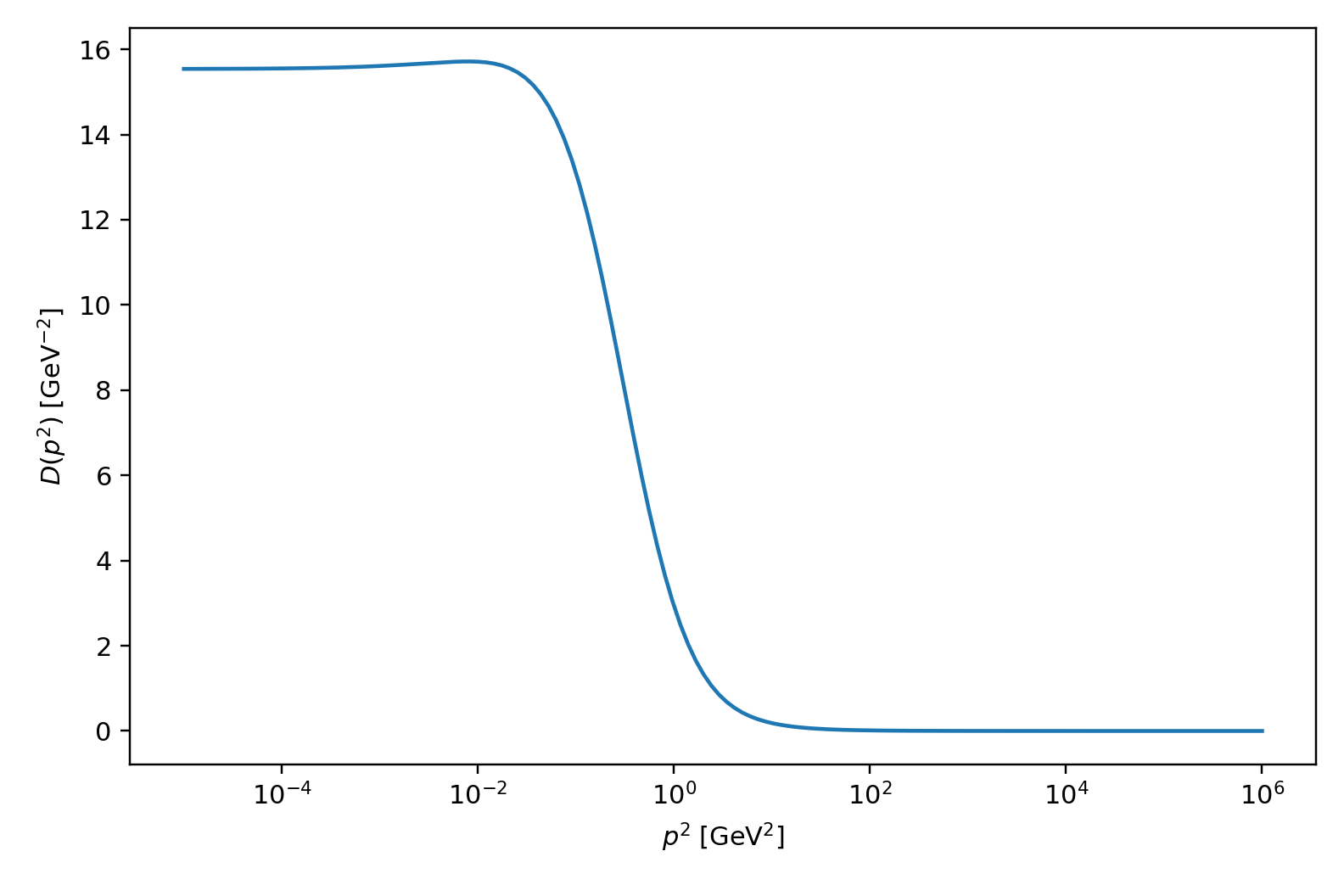}
 \caption{Scalar gluon propagator.}
\end{subfigure}
\caption{Direct baseline solution.  The dressing functions $G(p^2)$ and
$Z(p^2)$ are shown in the left panel.  The right panel shows
$D(p^2)=Z(p^2)/p^2$.  The infrared value is fixed by
$D(x_m)=15.54\,\GeV^{-2}$.}
\label{fig:direct-baseline}
\end{figure}

\begin{figure}
 \centering
 \includegraphics[width=0.67\textwidth]{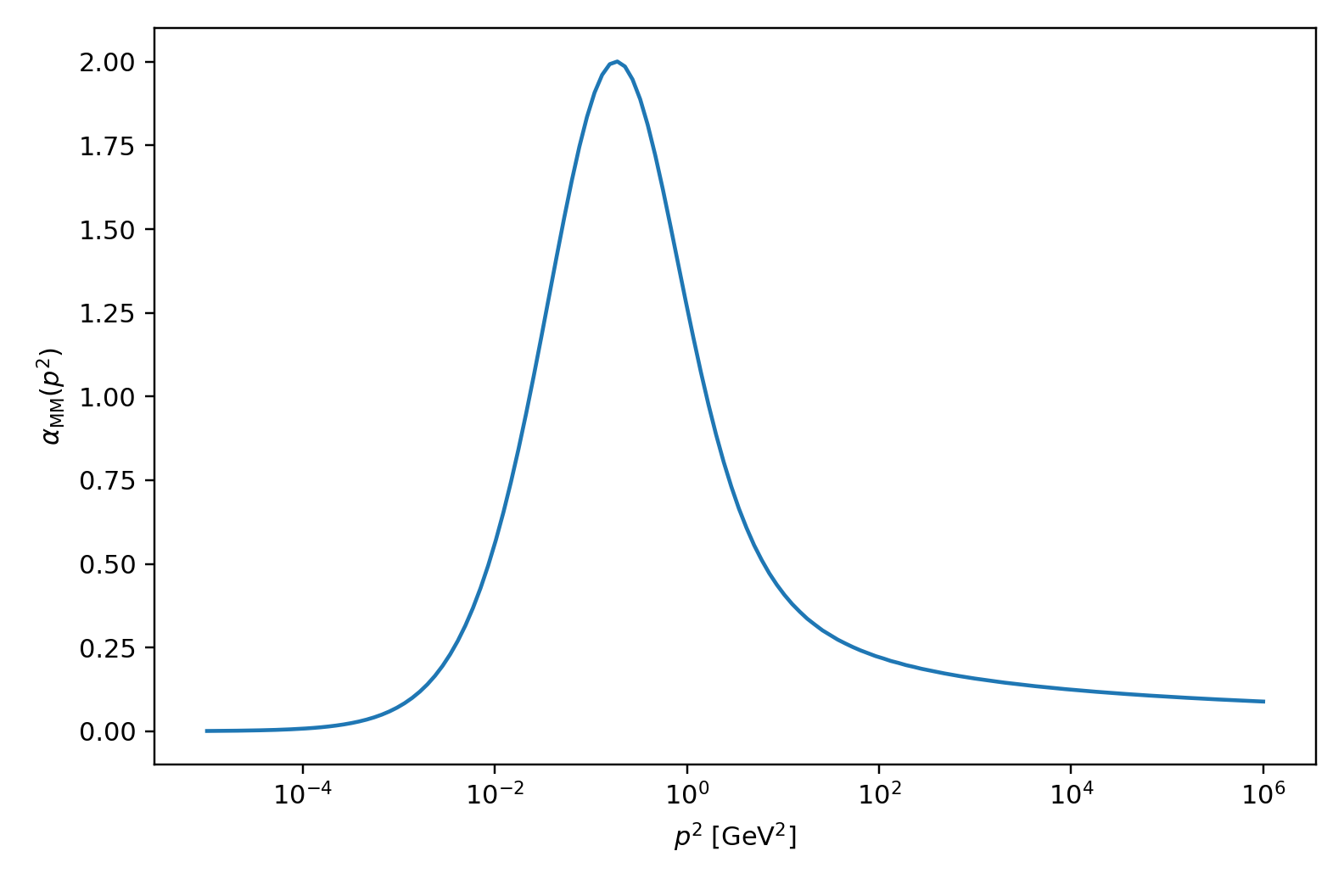}
 \caption{MiniMOM coupling obtained from the direct baseline solution.  The
 coupling vanishes in the infrared, has a maximum near
 $p^2=0.19\,\GeV^2$, and decreases logarithmically at large momenta.}
 \label{fig:minimom}
\end{figure}
In order to estimate the discretization error, the simulation was repeated
with three grids.  The errors in Table~\ref{tab:direct-convergence} are given
relative to the finest result.  For the baseline grid, the relative error is
below $5\times10^{-4}$ for the ghost dressing and about
$3.5\times10^{-3}$ for the gluon dressing.  The larger sensitivity of the
gluon equation will also be seen in the neural calculation.
\begin{table}
\centering
\begin{tabular}{c c c c c}
\toprule
calculation & $N_x$ & $N_y$ & $N_u$ & $(\varepsilon_2[G],\varepsilon_2[Z])$ \\
\midrule
coarse   & 100 & 90  & 24 & $(1.44\times10^{-3},\;1.06\times10^{-2})$\\
baseline & 140 & 140 & 36 & $(4.24\times10^{-4},\;3.52\times10^{-3})$\\
fine     & 190 & 200 & 52 & reference\\
\bottomrule
\end{tabular}
\caption{Direct solutions for three discretizations.  The relative errors are
calculated with respect to the fine solution on a common logarithmic momentum
grid.}
\label{tab:direct-convergence}
\end{table}
Starting from the analytic standard functions, the neural fixed-point
iteration converges without using the direct solution.  For the representative
baseline run, the relative errors are
\begin{equation}
 \varepsilon_2[G]=2.82\times10^{-3},
 \qquad
 \varepsilon_2[Z]=9.76\times10^{-3}.
 \label{eq:baseline-neural-errors}
\end{equation}
The largest pointwise differences are $0.71\%$ for the ghost dressing and
$2.07\%$ for the gluon dressing.  As in the direct refinement test, the gluon
is more sensitive to the numerical treatment.
\begin{figure}
\centering
\begin{subfigure}{0.48\textwidth}
 \includegraphics[width=\linewidth]{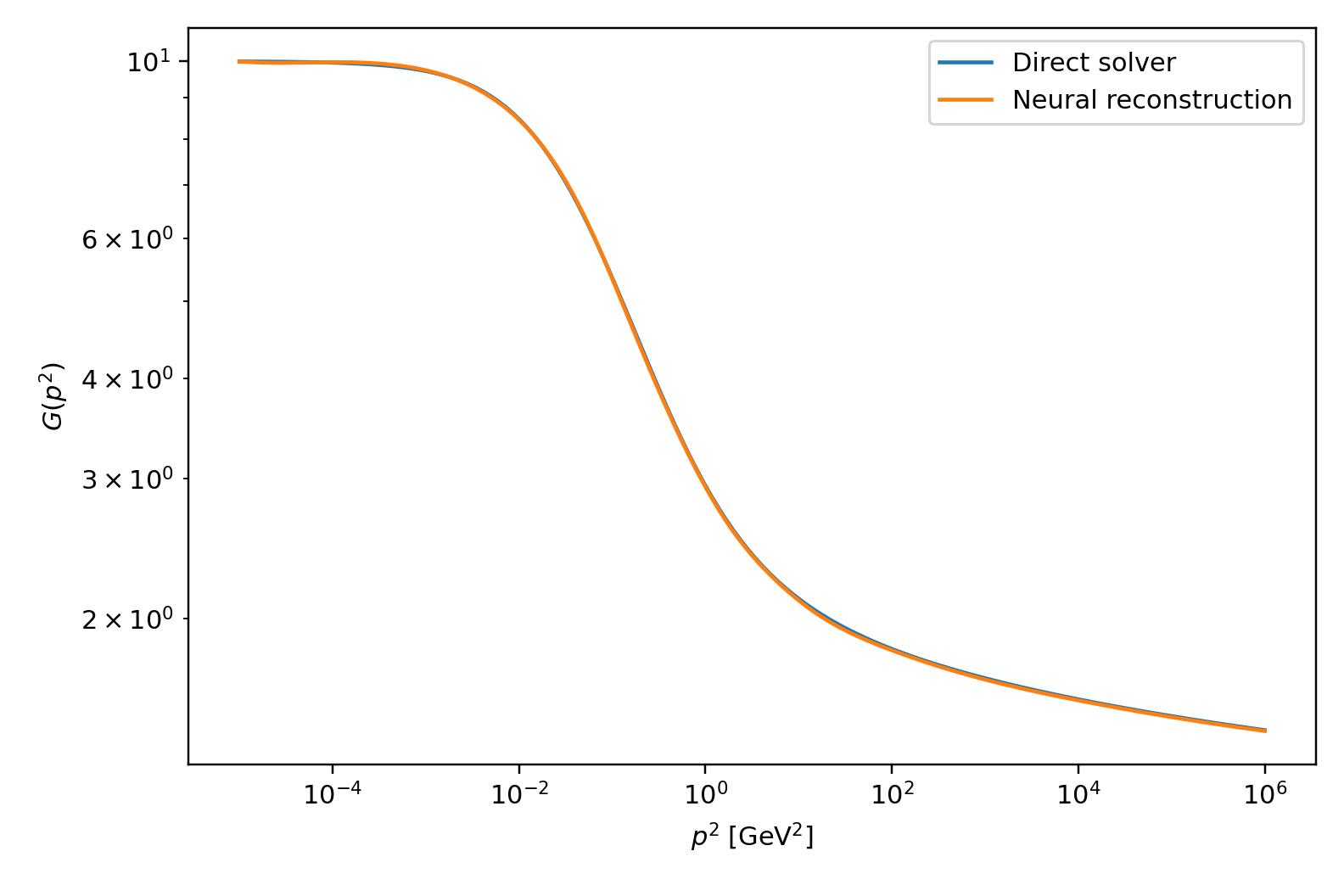}
 \caption{Ghost dressing.}
\end{subfigure}
\hfill
\begin{subfigure}{0.48\textwidth}
 \includegraphics[width=\linewidth]{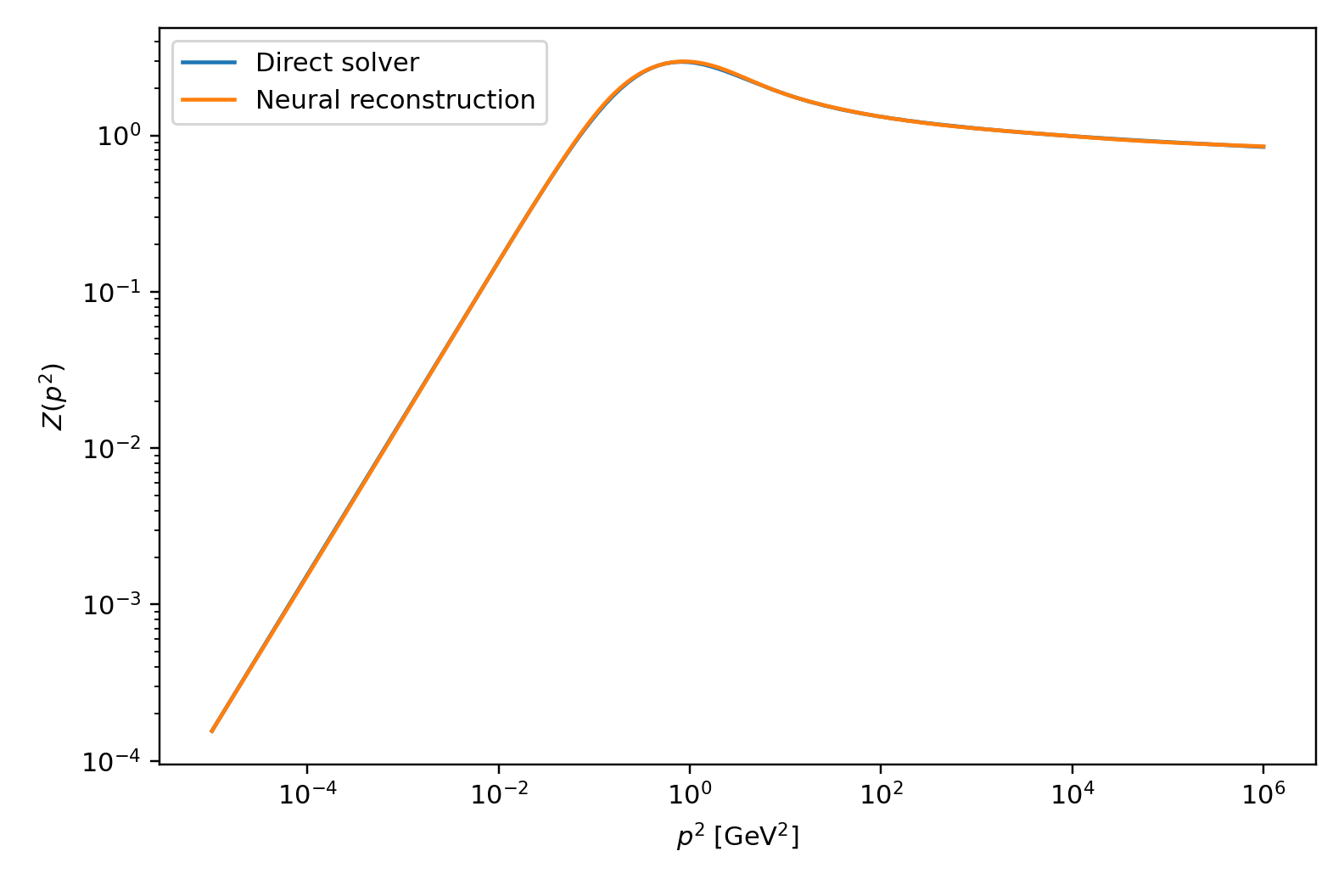}
 \caption{Gluon dressing.}
\end{subfigure}
\caption{Direct and neural solutions for the baseline truncation.  The direct
solution is not used during neural training.  The agreement extends over the
full momentum interval.}
\label{fig:neural-comparison}
\end{figure}

\begin{figure}
\centering
\begin{subfigure}{0.48\textwidth}
 \includegraphics[width=\linewidth]{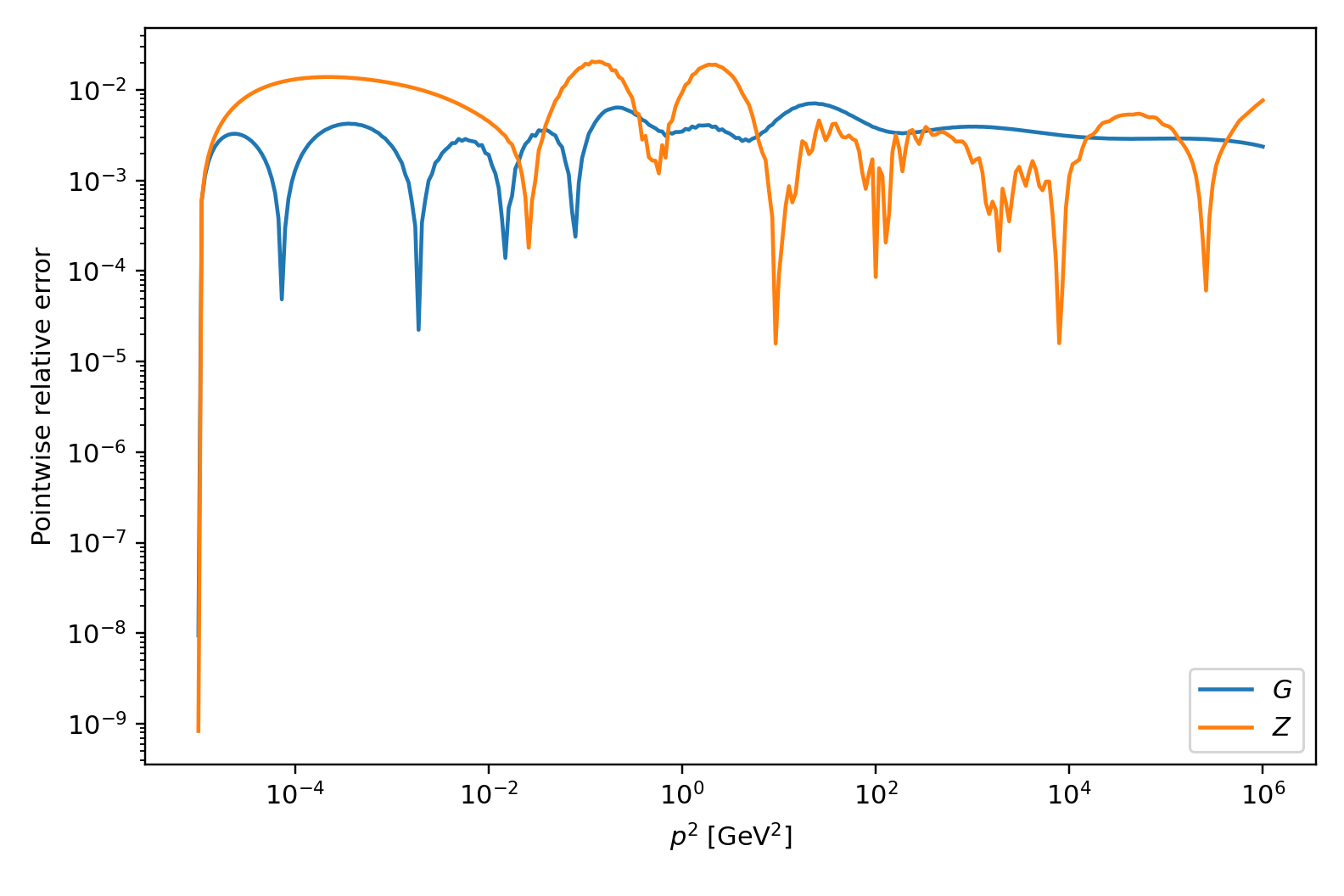}
 \caption{Pointwise relative differences.}
\end{subfigure}
\hfill
\begin{subfigure}{0.48\textwidth}
 \includegraphics[width=\linewidth]{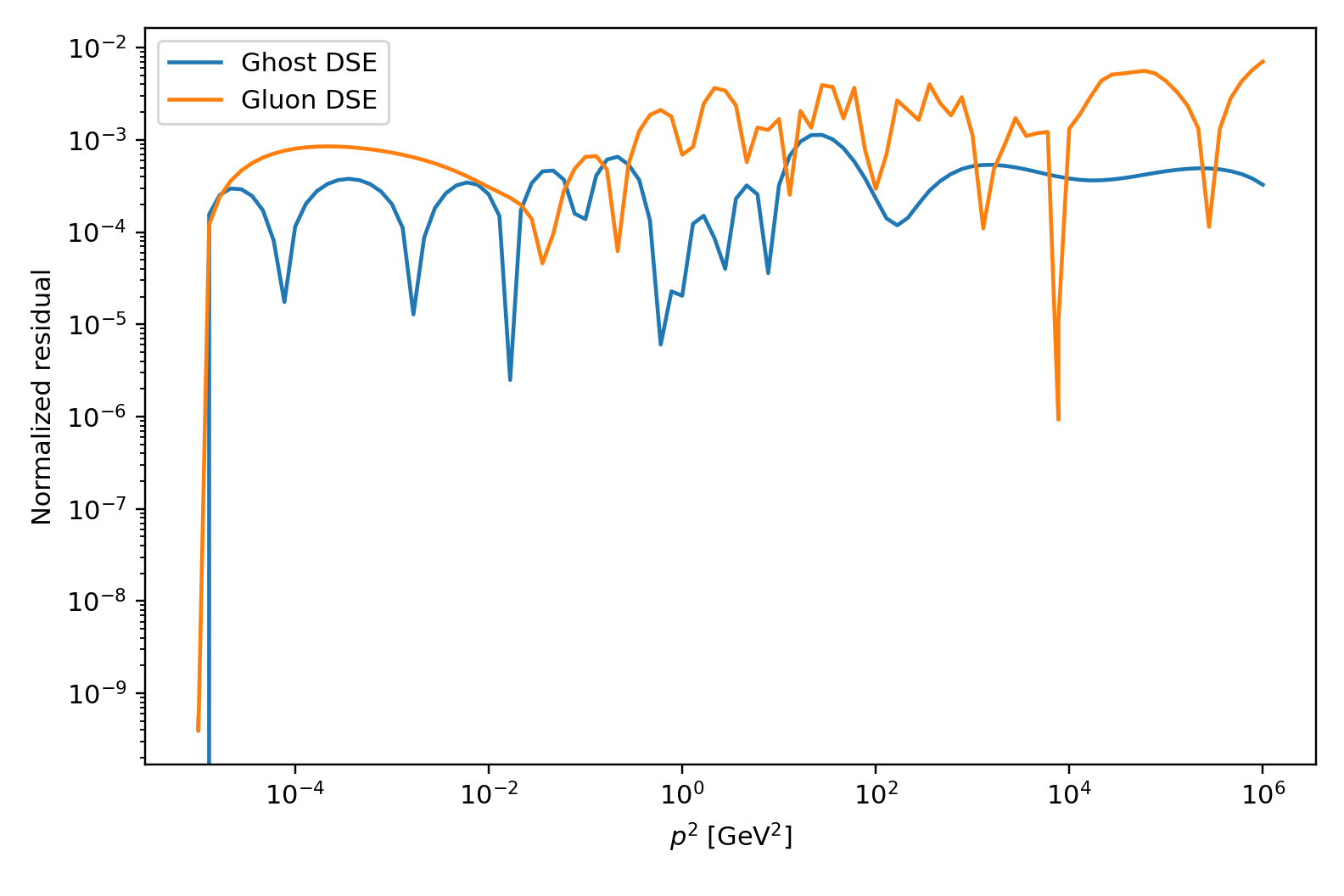}
 \caption{Residuals on the validation grid.}
\end{subfigure}
\caption{Numerical tests of the baseline neural solution.  The left panel
shows the difference from the direct result.  The right panel shows the ghost
and inverse-gluon-propagator residuals evaluated with a finer quadrature than
the one used in training.}
\label{fig:neural-diagnostics}
\end{figure}
On the independent validation grid, the largest residuals are
\begin{equation}
 \max_x|r_G(x)|=8.34\times10^{-4},
 \qquad
 \max_x|r_D(x)|=6.80\times10^{-3}.
 \label{eq:validation-residuals}
\end{equation}
The corresponding mean squared values are $8.8\times10^{-8}$ and
$5.0\times10^{-6}$.  The residual test is needed in addition to the comparison
with the direct curves.  Otherwise, agreement could result from interpolation
without an equally accurate solution of the integral equations.
Five parameter initializations were used for the width-$40$ network.  The last
layer was set to zero, so the initial dressing functions were the same, but
the hidden representations were different.  The resulting errors are
\begin{align}
 \varepsilon_2[G]&=(3.25\pm0.77)\times10^{-3},\\
 \varepsilon_2[Z]&=(9.48\pm0.29)\times10^{-3}.
 \label{eq:seed-errors}
\end{align}
Over the full momentum interval, the largest relative standard deviation is
$0.81\%$ for the ghost and $0.75\%$ for the gluon.  Three additional runs were
started from nonzero perturbations of the last layer and they converge to the
same branch as well.  Their ghost errors lie between
$3.78\times10^{-3}$ and $5.41\times10^{-3}$, and the gluon errors lie between
$7.15\times10^{-3}$ and $1.14\times10^{-2}$.

\begin{figure}
\centering
\begin{subfigure}{0.48\textwidth}
 \includegraphics[width=\linewidth]{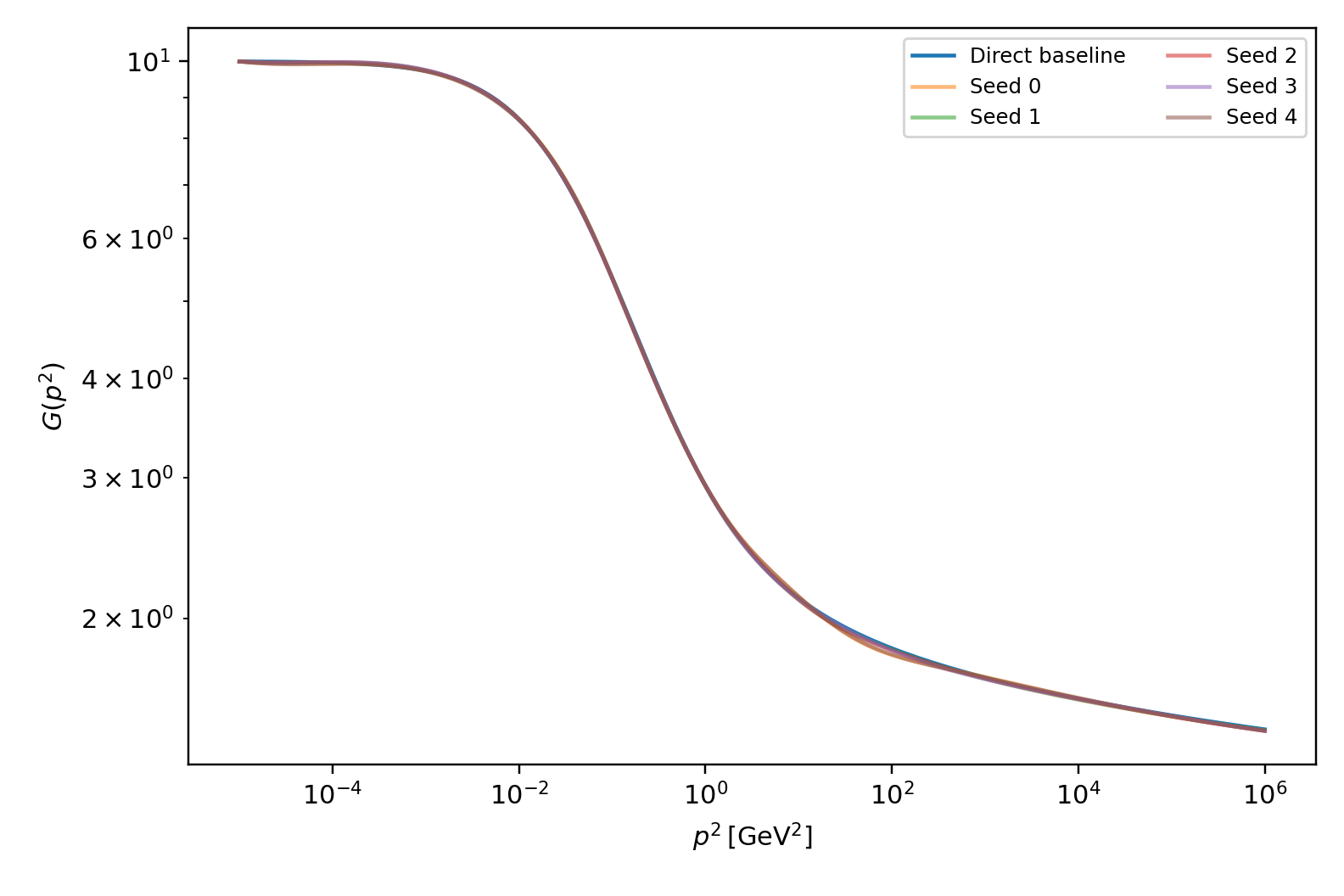}
 \caption{Ghost dressing.}
\end{subfigure}
\hfill
\begin{subfigure}{0.48\textwidth}
 \includegraphics[width=\linewidth]{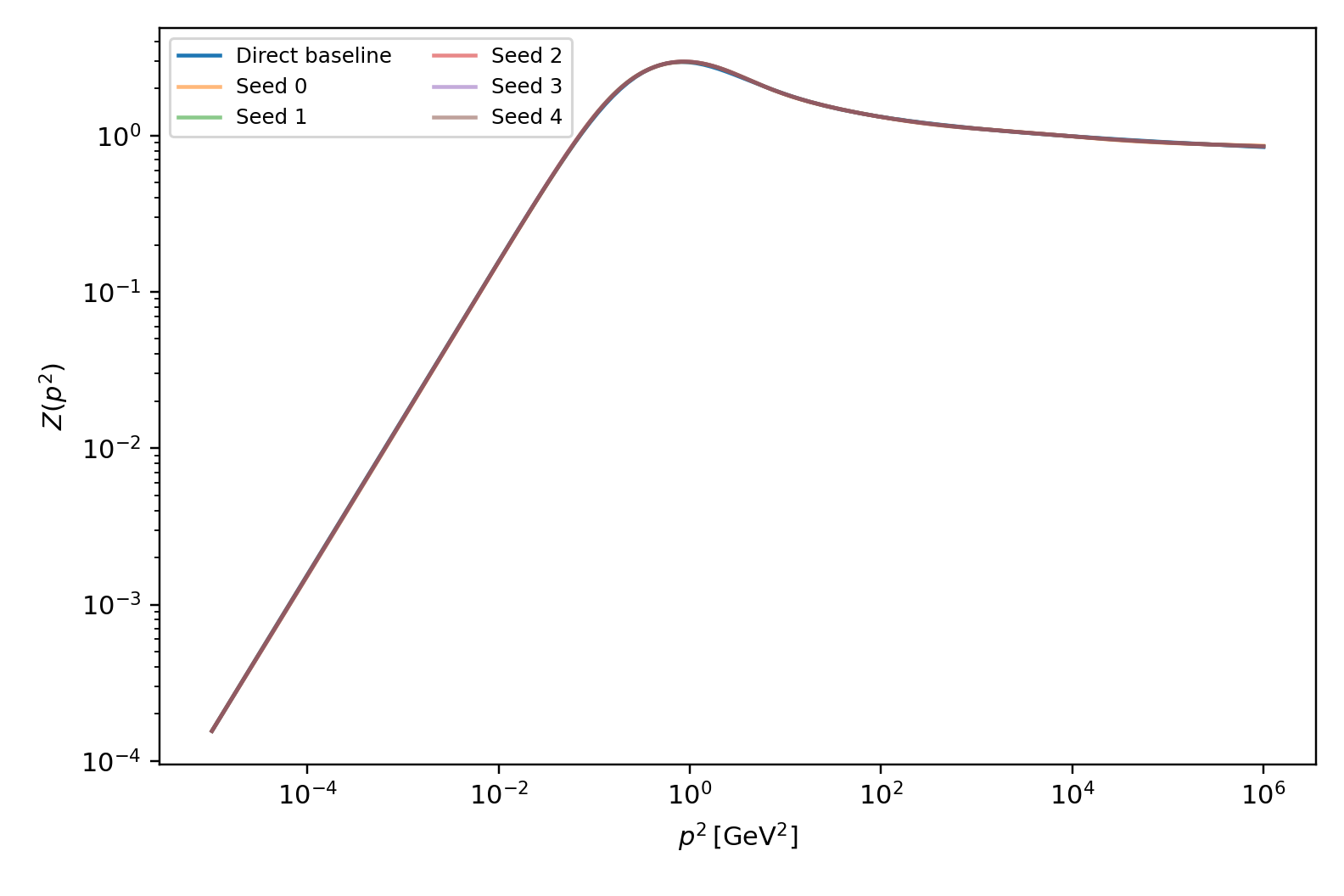}
 \caption{Gluon dressing.}
\end{subfigure}
\caption{Neural solutions for five parameter initializations.  The equations
and renormalization conditions are identical in all runs.}
\label{fig:seed-ensemble}
\end{figure}
The hidden-layer width was changed from $24$ to $64$.  The ghost error remains
close to $3\times10^{-3}$.  For the gluon it changes from
$9.06\times10^{-3}$ to $8.19\times10^{-3}$ and no monotonic dependence on the
width is found.  At this accuracy, the effects of optimization, quadrature,
and initialization are of comparable size.  The width scan therefore tests
stability but does not provide an asymptotic convergence law.
A clearer dependence is obtained for the integration grid.  Increasing the
radial--angular quadrature from $48\times12$ to $96\times24$ changes the gluon
error from $1.36\times10^{-2}$ to $8.36\times10^{-3}$.  The ghost error stays
between $2.8\times10^{-3}$ and $3.3\times10^{-3}$.  This confirms that the
gluon equation is the more demanding part of the coupled system.

\begin{figure}
\centering
\begin{subfigure}{0.48\textwidth}
 \includegraphics[width=\linewidth]{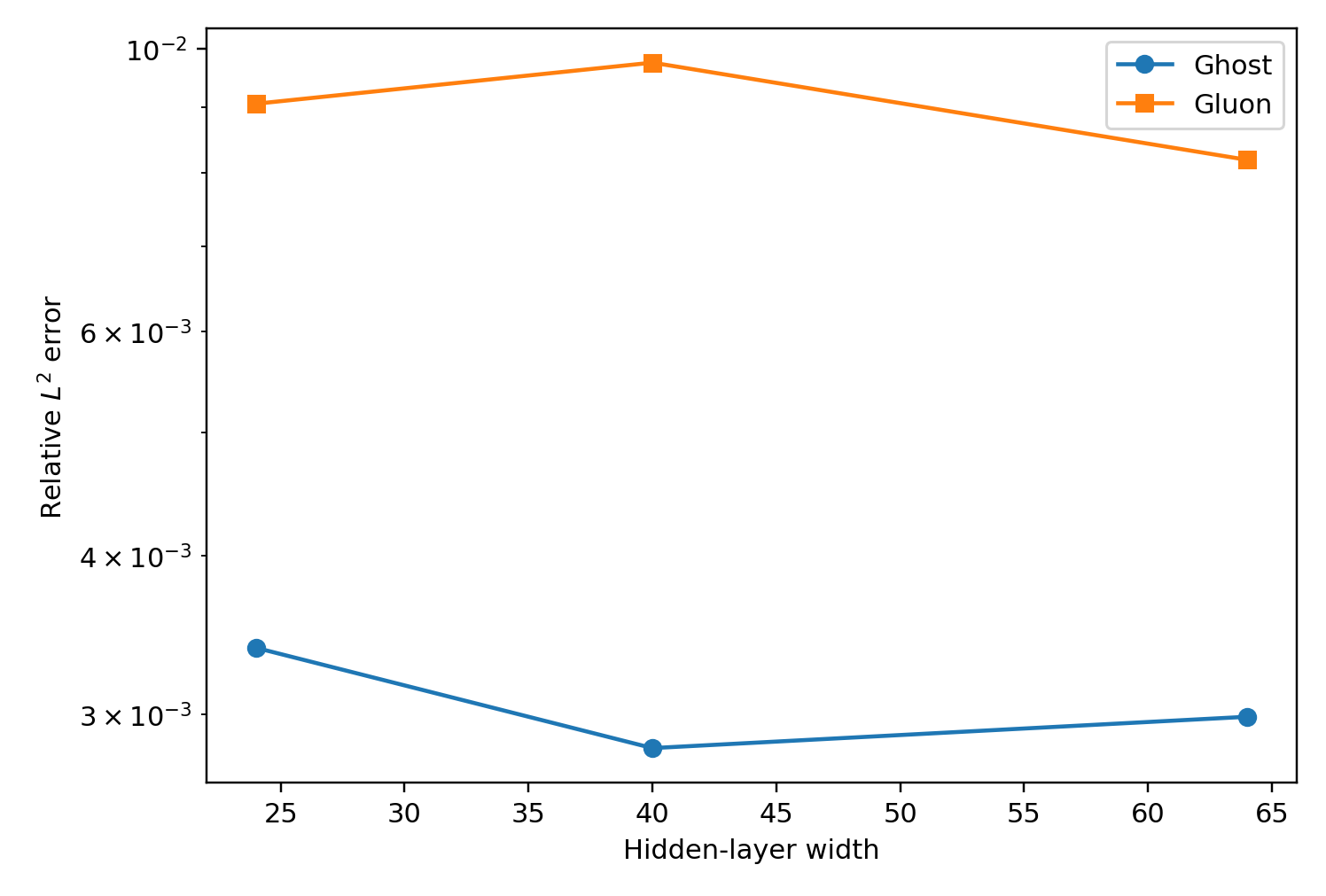}
 \caption{Variation of the network width.}
\end{subfigure}
\hfill
\begin{subfigure}{0.48\textwidth}
 \includegraphics[width=\linewidth]{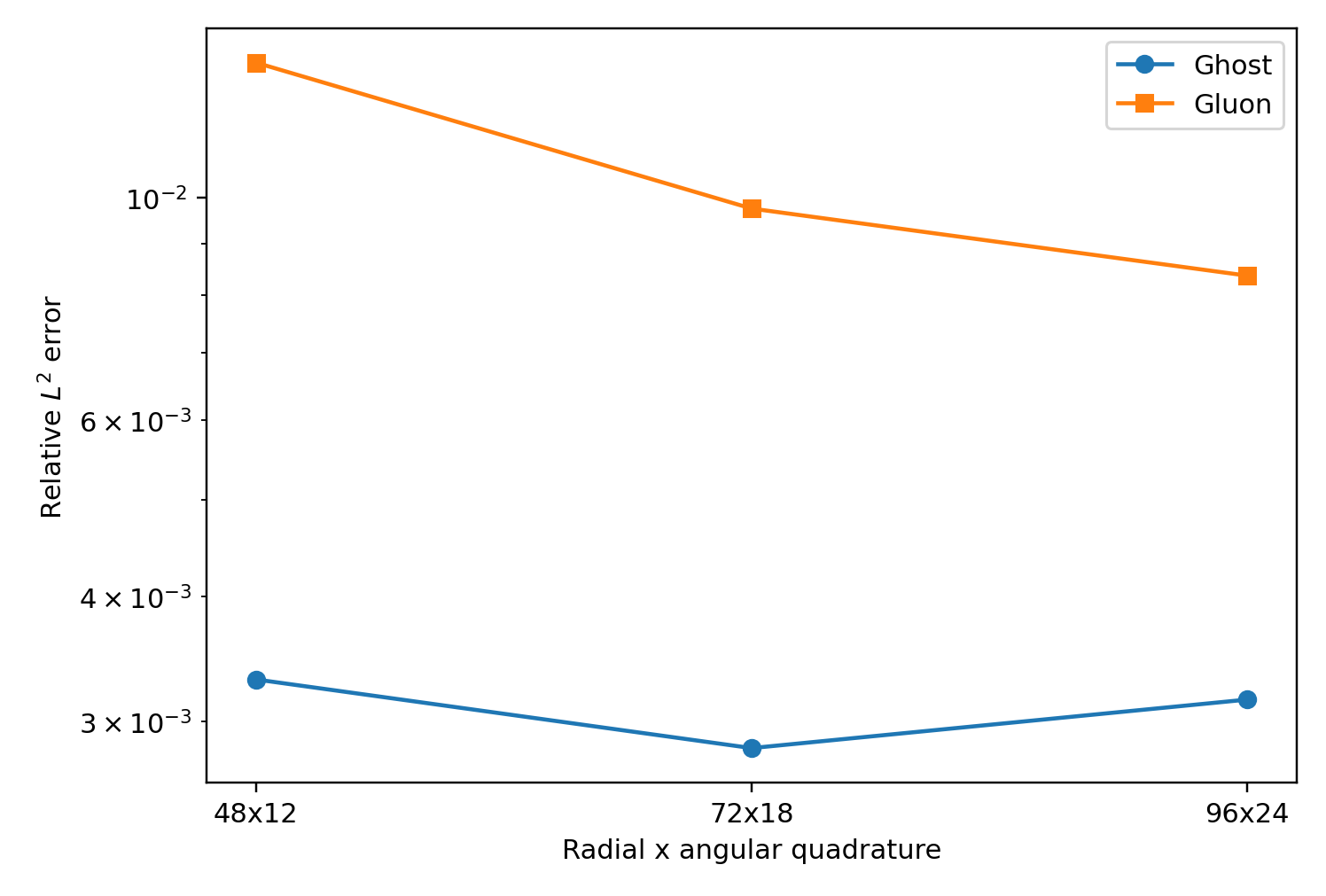}
 \caption{Variation of the training quadrature.}
\end{subfigure}
\caption{Relative neural errors for different network widths and integration
grids.  The gluon solution changes more under quadrature refinement than under
the variation of the width.}
\label{fig:capacity-quadrature}
\end{figure}

The largest changes are attained when the three-gluon-vertex model is varied.
The bare and suppression-only models both lead to positive Euclidean
solutions.  Compared with the baseline calculation, the gluon dressing
changes by about $33$--$36\%$ in relative $L^2$ norm.  The maxima of the gluon
dressing and of the MiniMOM coupling are given in
Table~\ref{tab:vertex-results}.  For each of the three truncations, the neural
solution remains below the percent level relative to the corresponding direct
solution.

\begin{figure}
\centering
\begin{subfigure}{0.48\textwidth}
 \includegraphics[width=\linewidth]{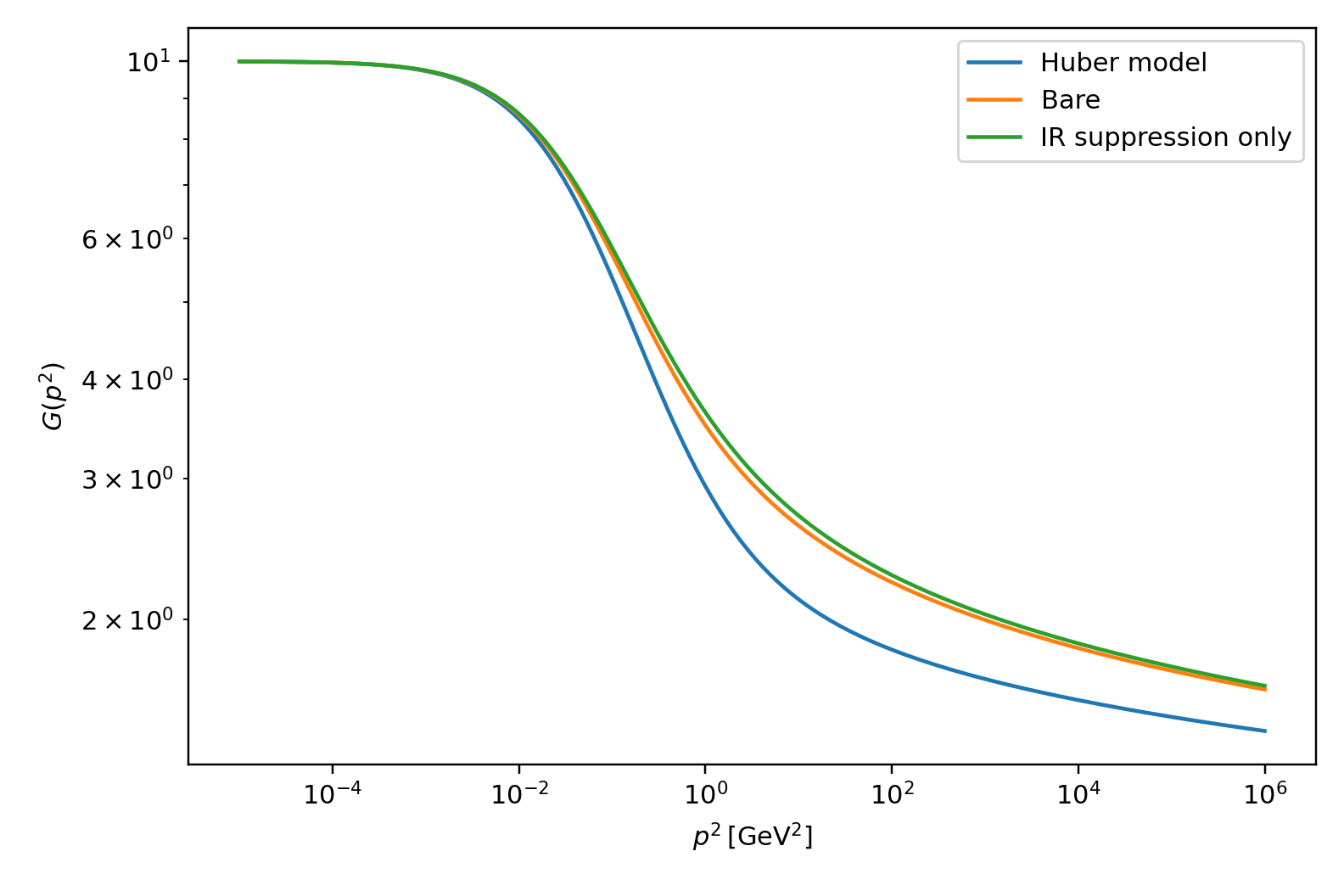}
 \caption{Ghost dressing.}
\end{subfigure}
\hfill
\begin{subfigure}{0.48\textwidth}
 \includegraphics[width=\linewidth]{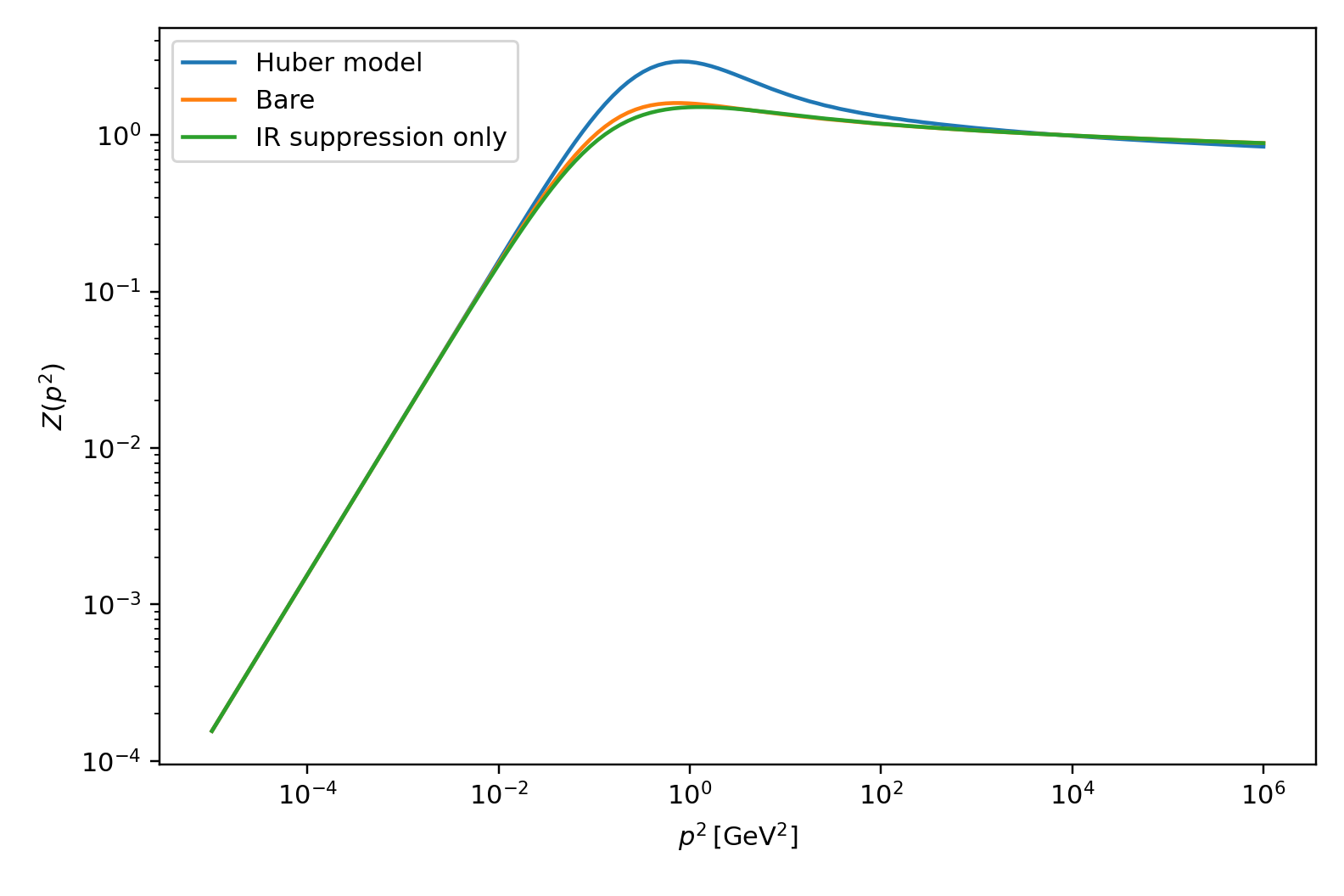}
 \caption{Gluon dressing.}
\end{subfigure}
\caption{Direct solutions for three models of the three-gluon vertex.  The
same renormalization conditions are used in all cases.  The main change occurs
in the gluon dressing around its maximum.}
\label{fig:vertex-dependence}
\end{figure}

\begin{table}
\centering
\begin{tabular}{l c c c c}
\toprule
vertex model & $Z_{\max}$ & $\alpha_{\MM}^{\max}$ & $\varepsilon_2[G_\theta]$ & $\varepsilon_2[Z_\theta]$\\
\midrule
baseline $G/Z$ with IR suppression & 2.954 & 2.001 & $2.82\times10^{-3}$ & $9.76\times10^{-3}$\\
bare & 1.602 & 1.643 & $3.16\times10^{-3}$ & $5.14\times10^{-3}$\\
IR suppression only & 1.510 & 1.536 & $3.57\times10^{-3}$ & $2.07\times10^{-3}$\\
\bottomrule
\end{tabular}
\caption{Results for the three convergent vertex models.  The neural errors
are calculated with the direct solution of the same truncation as reference.}
\label{tab:vertex-results}
\end{table}
The model without infrared suppression does not converge to the same positive
branch.  Before the fixed-point iteration breaks down, the maximum of the
gluon dressing grows to about $9.5$ and the MiniMOM coupling reaches about
$4.69$.  The instability occurs already in the direct calculation.  It is
therefore caused by the truncated equations and not by the neural
representation.
For the computations considered here, the vertex dependence is much larger
than the direct discretization error or the neural reconstruction error.  This
observation is important for the interpretation of a numerically accurate
neural result.  A more accurate solution of a fixed operator does not reduce
the uncertainty caused by the modeled vertex.

The decoupling solutions of the functional equations are selected by an
infrared boundary condition~\cite{FischerMaasPawlowski2009,Huber2020Review}.
We use
\begin{equation}
 G(x_m)=5,\;10,\;15,\;20
\end{equation}
while $D(x_m)=15.54\,\GeV^{-2}$ is kept fixed.  The direct iteration converges
for all four values.  The main change in the gluon dressing occurs around its
maximum.  The maximum of the MiniMOM coupling increases from $1.16$ to $2.58$.

\begin{figure}
\centering
\begin{subfigure}{0.48\textwidth}
 \includegraphics[width=\linewidth]{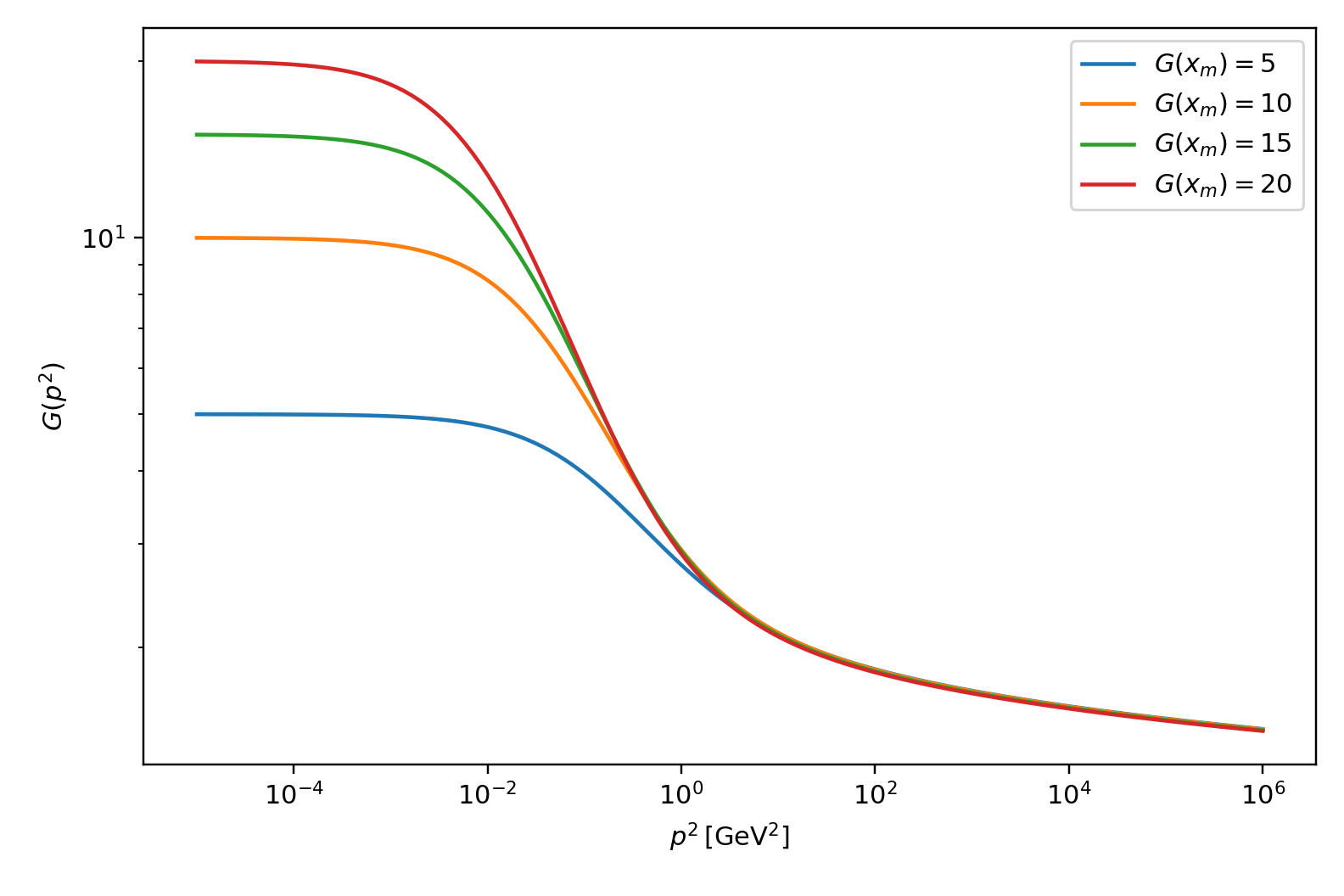}
 \caption{Ghost dressing.}
\end{subfigure}
\hfill
\begin{subfigure}{0.48\textwidth}
 \includegraphics[width=\linewidth]{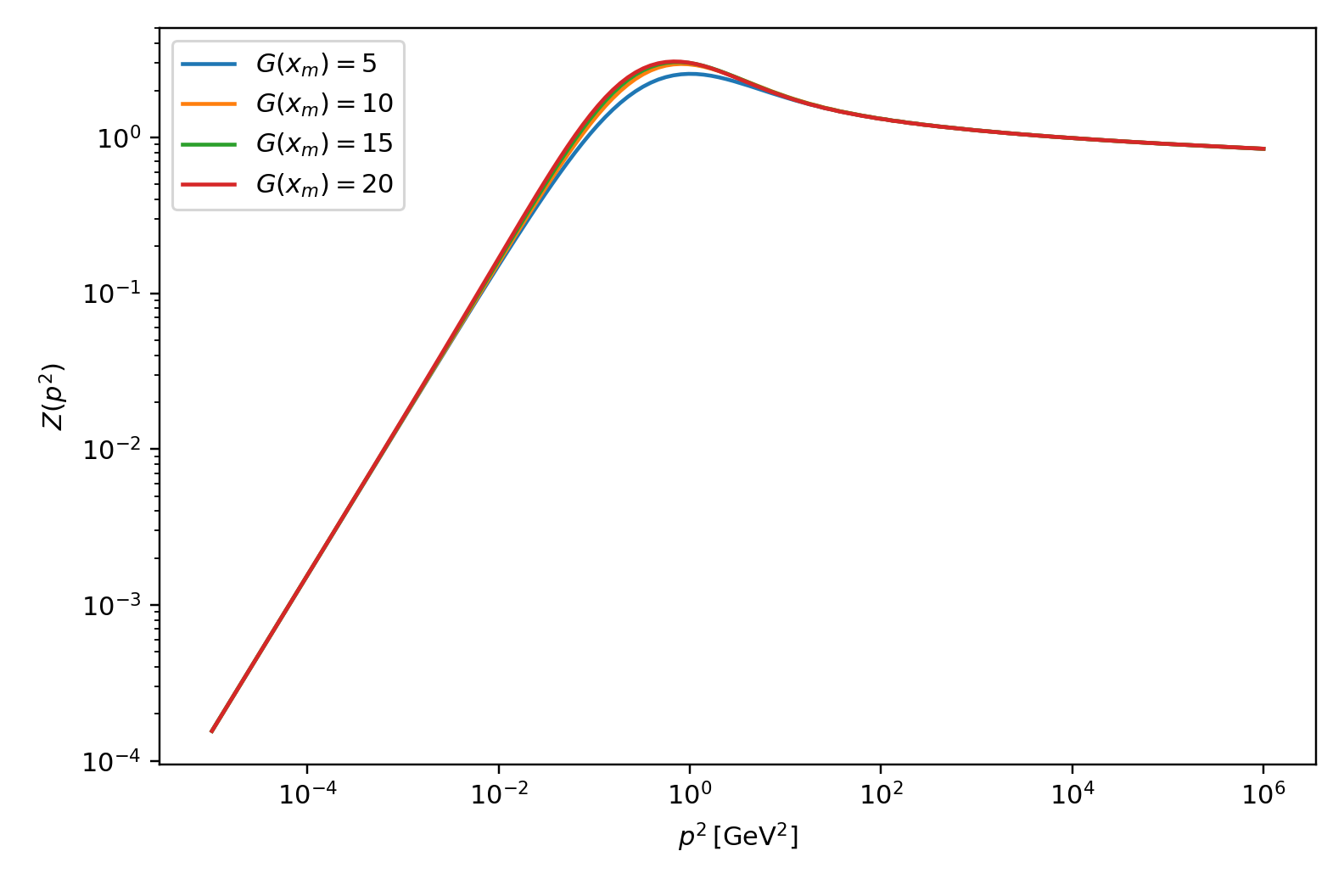}
 \caption{Gluon dressing.}
\end{subfigure}
\caption{Direct solutions for four infrared ghost conditions.  The infrared
gluon propagator is kept fixed.  The solutions have the same ultraviolet
normalization but differ in the transition region.}
\label{fig:infrared-family}
\end{figure}
The neural calculation was also performed for $G(x_m)=5$, $15$, and $20$.  The
relative ghost errors are $2.47\times10^{-3}$, $3.21\times10^{-3}$, and
$4.26\times10^{-3}$.  The corresponding gluon errors are
$7.76\times10^{-3}$, $7.47\times10^{-3}$, and $1.09\times10^{-2}$.  Hence,
the neural iteration is not restricted to the baseline boundary condition.
It follows the different members of the decoupling family.

For the ultraviolet (UV) analysis, the external momentum interval is extended to
$p^2=10^{10}\,\GeV^2$ and the radial cutoff to $10^{12}\,\GeV^2$.  In the
highest fit interval, the direct solution gives
\begin{equation}
 \delta_{\rm dir}=-0.2744,
 \qquad
 \gamma_{\rm dir}=-0.4467.
 \label{eq:direct-uv-exponents}
\end{equation}
The distinct values differ from $-9/44$ and $-13/22$.  The reduced one-loop
truncation and the effective vertex therefore do not reproduce the two
anomalous dimensions separately.  Their combination is
\begin{equation}
 2\delta_{\rm dir}+\gamma_{\rm dir}=-0.9955,
 \label{eq:direct-uv-combination}
\end{equation}
which is close to the one-loop value $-1$ of the MiniMOM coupling.

\begin{figure}
 \centering
 \includegraphics[width=0.69\textwidth]{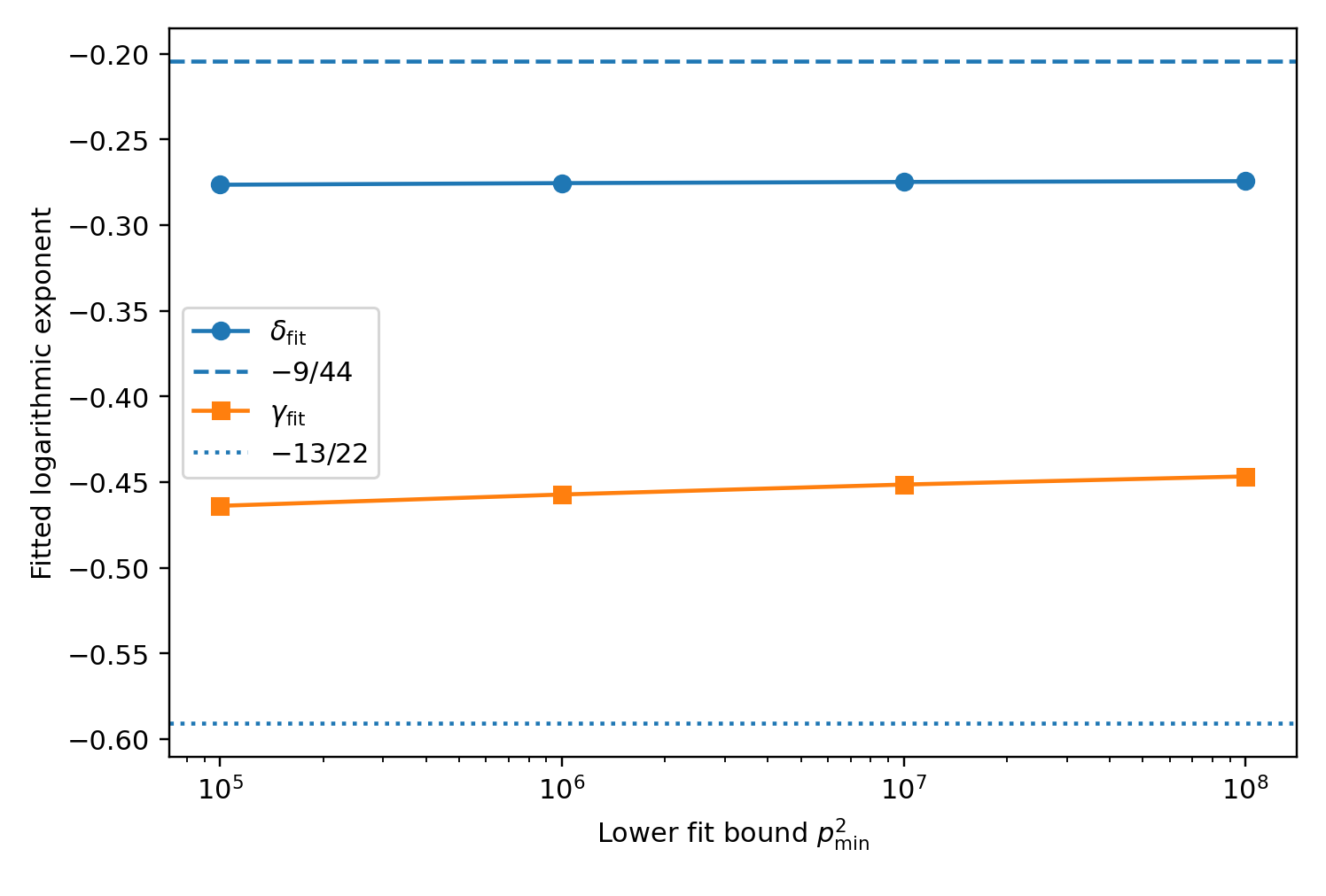}
 \caption{Ultraviolet exponents of the extended direct solution for different
 lower boundaries of the fit interval.  The horizontal lines show
 $\delta=-9/44$ and $\gamma=-13/22$.}
 \label{fig:uv-exponents}
\end{figure}
For the extended neural solution, the relative errors are
\begin{equation}
 \varepsilon_2[G]=7.35\times10^{-3},
 \qquad
 \varepsilon_2[Z]=1.80\times10^{-2}.
\end{equation}
The combination extracted from the highest interval is
\begin{equation}
 2\delta_\theta+\gamma_\theta=-1.012.
 \label{eq:neural-uv-combination}
\end{equation}
Thus, our neural simulation reproduces both features of the direct result:
the correct MiniMOM combination and the displaced individual exponents.  It
should be noted that this is the expected behavior of a solver which uses only
the truncated equations.  The missing UV structure has to be
improved in the truncation and cannot be supplied by the numerical method.

Finally, the Schwinger function is evaluated by a direct cosine integration.  The
cutoffs $p_{\max}=100$, 300, and $1000\,\GeV$ are used.  The first zero is
stable under this variation.  Results for $p_{\max}=1000\,\GeV$ are listed in
Table~\ref{tab:schwinger}.

\begin{figure}
 \centering
 \includegraphics[width=0.72\textwidth]{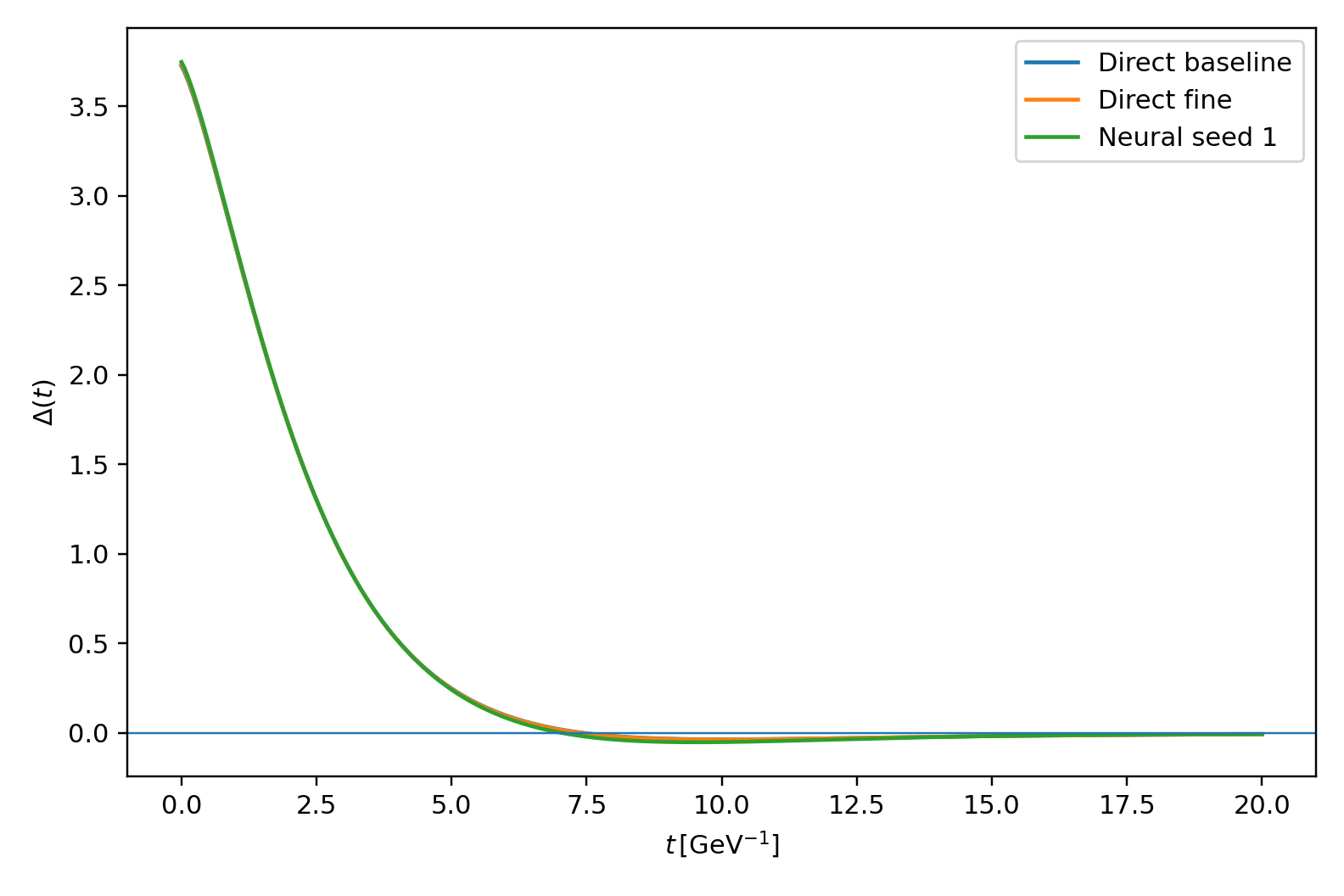}
 \caption{Gluon Schwinger function from the baseline direct, fine direct, and
 neural propagators.  All three functions become negative.  The first zero of
 the neural result occurs at a smaller Euclidean time.}
 \label{fig:schwinger}
\end{figure}

\begin{table}
\centering
\begin{tabular}{l c c}
\toprule
solution & first zero $t_0$ [$\GeV^{-1}$] & minimum of $\Delta(t)$\\
\midrule
direct baseline & 7.459 & $-3.53\times10^{-2}$\\
direct fine & 7.408 & $-3.60\times10^{-2}$\\
neural, representative seed & 7.054 & $-5.11\times10^{-2}$\\
\bottomrule
\end{tabular}
\caption{First zero and minimum of the gluon Schwinger function at
$p_{\max}=1000\,\GeV$.}
\label{tab:schwinger}
\end{table}
The two direct discretizations give first zeros which differ by less than one
percent.  The neural zero is shifted by about five percent relative to the
fine direct result.  Nevertheless, the negative part is present in all cases.
Thus, the violation of reflection positivity is reproduced qualitatively.
The position and depth of the negative region are more sensitive than the
Euclidean dressing functions themselves. 

\section{Discussion}
\label{sec:discussion}

The results enable the different sources of uncertainty to be compared.  The
first one is the discretization of the direct integral equations.  For the
reference setup grid it is below the per-mille level for the ghost and at a few
per mille for the gluon.  The second one is our neural solution of these
discretized equations.  Its error is somewhat larger, but remains at the
per-mille to percent level.  The tests with different initializations,
network widths, and integration grids indicate that the same fixed point is
obtained.  The residuals on the independent grid support this conclusion.
A much larger effect is caused by the three-gluon vertex.  Replacing the
initial model by a bare or suppression-only dressing changes the gluon
solution by more than thirty percent.  Our neural solution remains accurate
for each of these choices.  For the present truncation, the sizes of the three
effects can therefore be summarized as
\begin{equation}
 \text{direct discretization error}
 \lesssim
 \text{neural reconstruction error}
 \ll
 \text{three-gluon-vertex dependence}.
 \label{eq:error-hierarchy}
\end{equation}
This comparison also shows why the numerical and truncation errors should be
kept separate.  Increasing the network size cannot compensate for missing or
modeled correlation functions.
The UV calculation provides an example of this point.  The direct
truncation does not yield the correct anomalous dimensions of the two
propagators separately.  It does, however, reproduce their MiniMOM combination
with good accuracy.  Our neural simulation gives the same result.  This is
not a shortcoming of our neural solver, it reflects the perturbative content
of the equations which are solved. 

The Schwinger function is more sensitive to the propagator than the pointwise
comparison on the Euclidean axis.  The direct and neural propagators differ by
only a few percent, whereas the first zero changes by about five percent and
the minimum changes more strongly.  Oscillatory transforms can therefore
require a higher accuracy than the one needed for the original dressing
functions.  The negative part itself is stable.  In this respect, the result
is consistent with the observation that positivity properties should be
checked after the solution and not imposed in a way which removes the signal
of their violation~\cite{Terin2026JHEP}.
Several restrictions of the calculation should be kept in mind.  The
ghost--gluon vertex is bare.  The three-gluon vertex contains only one modeled
scalar dressing.  The four-gluon vertex and the two-loop diagrams of the gluon
equation are not included.  The mass counterterm removes the leading cutoff
dependence, but its treatment is part of the truncation.  Furthermore, only
Euclidean pure YM theory is considered.  The results are therefore a
study of the numerical solution of a fixed propagator truncation and not a
new state-of-the-art calculation of YM correlation functions.

The vertex test indicates the most useful next step.  A neural representation
of the three-gluon vertex could be coupled to its DS or 3PI
equation.  This would reduce the model input in the propagator system.  It
would also provide a more demanding test of the neural procedure because a
three-point function depends on several momentum variables.  A
parameter-dependent network for the infrared family is another possible
extension.  Complex momenta and Minkowski space require additional work.  In
that case, sign-indefinite spectral functions and complex singularities have
to be enabled from the beginning.

\section{Conclusions}
\label{sec:conclusions}

The coupled ghost and gluon DSEs were solved with our
neural representation.  The calculation was performed in four-dimensional
YM theory in Landau gauge.  A one-loop propagator truncation with a
modeled three-gluon vertex was used.  The same renormalized equations were
also solved by a conventional fixed-point iteration.

It is important to remark that our neural computation does not use converged propagator data.  Targets from
the equations are employed for a fixed-point preconditioning, and the
renormalized residuals are minimized afterwards.  For the baseline system,
the relative errors are $2.8\times10^{-3}$ for the ghost dressing and
$9.8\times10^{-3}$ for the gluon dressing.  Similar results are attained for
different initializations, network widths, integration grids, and infrared
boundary conditions.  The residuals evaluated with a finer quadrature show
that the agreement is not restricted to the training points.
It turns out that the model for the three-gluon vertex is more important than
the remaining numerical error.  The alternative vertex models change the
gluon dressing by more than thirty percent, while our neural solution of each
fixed truncation stays below the percent level.  The UV test gives a
similar lesson.  The MiniMOM combination of anomalous dimensions is obtained,
but the distinct exponents are not.  Our neural result follows this behavior
of the truncated equations.  The Schwinger function becomes negative for the
direct and neural propagators, although the position of its first zero is more
sensitive to small propagator differences.
Neural representations can therefore be used for coupled and renormalized
functional equations with an accuracy comparable to conventional numerical
methods.  Their usefulness does not remove the usual truncation problem.  For
the system considered here, an improvement of the vertex sector is more
important than a further increase of the propagator-network size.  A
dynamical neural calculation of the three-gluon vertex is a natural next step.

\appendix

\section{Summary of neural convergence tests}
\label{app:neural-tests}

\begin{table}[h]
\centering
\begin{tabular}{c c c}
\toprule
hidden width & $\varepsilon_2[G]$ & $\varepsilon_2[Z]$\\
\midrule
24 & $3.39\times10^{-3}$ & $9.06\times10^{-3}$\\
40 & $2.82\times10^{-3}$ & $9.76\times10^{-3}$\\
64 & $2.99\times10^{-3}$ & $8.19\times10^{-3}$\\
\bottomrule
\end{tabular}
\caption{Network-width scan at fixed depth and quadrature.}
\label{tab:width-scan}
\end{table}

\begin{table}[h]
\centering
\begin{tabular}{c c c}
\toprule
radial--angular quadrature & $\varepsilon_2[G]$ & $\varepsilon_2[Z]$\\
\midrule
$48\times12$ & $3.31\times10^{-3}$ & $1.36\times10^{-2}$\\
$72\times18$ & $2.82\times10^{-3}$ & $9.76\times10^{-3}$\\
$96\times24$ & $3.16\times10^{-3}$ & $8.36\times10^{-3}$\\
\bottomrule
\end{tabular}
\caption{Dependence of the neural reconstruction on the quadrature used during training.  All residuals are subsequently evaluated on a common finer validation quadrature.}
\label{tab:quadrature-scan}
\end{table}

\begin{table}[h]
\centering
\begin{tabular}{c c c c}
\toprule
$G(x_m)$ & $Z_{\max}$ & $\alpha_{\MM}^{\max}$ & $(\varepsilon_2[G],\varepsilon_2[Z])$\\
\midrule
5  & 2.553 & 1.164 & $(2.47\times10^{-3},\;7.76\times10^{-3})$\\
10 & 2.954 & 2.001 & $(2.82\times10^{-3},\;9.76\times10^{-3})$\\
15 & 3.036 & 2.376 & $(3.21\times10^{-3},\;7.47\times10^{-3})$\\
20 & 3.065 & 2.585 & $(4.26\times10^{-3},\;1.09\times10^{-2})$\\
\bottomrule
\end{tabular}
\caption{Direct observables and neural errors across the infrared family.  The baseline row is included for comparison.}
\label{tab:infrared-family}
\end{table}
\newpage

\section{Numerical parameters}
\label{app:parameters}

\begin{table}[h]
\centering
\begin{tabular}{l l}
\toprule
quantity & baseline value\\
\midrule
color number & $N_c=3$\\
subtraction-point coupling & $\alpha(x_s)=0.05$\\
infrared ghost condition & $G(x_m)=10$\\
infrared gluon condition & $D(x_m)=15.54\,\GeV^{-2}$\\
infrared point & $x_m=10^{-5}\,\GeV^2$\\
subtraction point & $x_s=7720\,\GeV^2$\\
three-gluon damping scale & $\Lambda_s^2=1.54\,\GeV^2$\\
direct external/radial/angular points & $140/140/36$\\
neural training external/radial/angular points & $80/72/18$\\
neural validation external/radial/angular points & $120/112/30$\\
hidden layers and width & $2\times40$\\
outer neural Picard steps & 70\\
Adam projection steps per outer iteration & 50\\
residual fine-tuning steps & 60\\
\bottomrule
\end{tabular}
\caption{Baseline physical and numerical parameters.}
\label{tab:parameters}
\end{table}

\bibliographystyle{unsrtnat}
\bibliography{refs}

\end{document}